\documentclass[lettersize,journal]{IEEEtran}
\usepackage{amsmath,amsfonts, amsthm}
\usepackage{algorithmic}
\usepackage{algorithm}
\usepackage{array}
\usepackage[caption=false,font=normalsize,labelfont=sf,textfont=sf]{subfig}
\usepackage{textcomp}
\usepackage{stfloats}
\usepackage{url}
\usepackage{verbatim}
\usepackage{graphicx}
\usepackage{xcolor}
\usepackage{cite}
\usepackage{bm}
\usepackage{enumitem}
\usepackage{multirow}
\usepackage{mathtools}
\usepackage{cellspace} %
\definecolor{color0}{rgb}{0.12156862745098,0.466666666666667,0.705882352941177}
\definecolor{color1}{rgb}{1,0.498039215686275,0.0549019607843137}
\definecolor{color2}{rgb}{0.172549019607843,0.627450980392157,0.172549019607843}
\definecolor{color3}{rgb}{0.83921568627451,0.152941176470588,0.156862745098039}
\definecolor{color4}{rgb}{0.580392156862745,0.403921568627451,0.741176470588235}

\newtheorem{lemma}{Lemma}   
\newtheorem{definition}{Definition}   
\newtheorem{remark}{Remark}   
\newtheorem{prop}{Proposition}    
\newtheorem{theorem}{Theorem}  
\newtheorem{assum}{Assumption} 
\newtheorem*{corol*}{Corollary}

\newcommand{\x}{\bm x}
\newcommand{\esth}{^{(h)}}
\newcommand{\estmh}{^{(-h)}}

\newcommand{\xhat}{\hat{\bm x}}

\newcommand{\bary}{\bar{\bm y}}

\newcommand{\xstar}{\bm{x}^*}
\newcommand{\G}{\bm G}

\newcommand{\M}{\bm M}
\newcommand{\F}{\bm F}
\newcommand{\Qt}{\bm Q^T}
\newcommand{\Q}{\bm Q}

\newcommand{\R}{\bm R}
\newcommand{\W}{\bm W}
\newcommand{\V}{\bm V}
\newcommand{\Gset}{\mathcal{G}}
\newcommand{\Vset}{\mathcal{V}}
\newcommand{\Eset}{\mathcal{E}}
\newcommand{\I}{\bm I}
\newcommand{\Hb}{\bm H}
\newcommand{\y}{\bm y}
\newcommand{\gh}{\bm g^h}

\newcommand{\1}{\bm 1}

\newcommand{\nl}{\left| \left|}
\newcommand{\nr}{\right| \right|}
\newcommand{\nrq}{\right| \right|^2}

\newcommand{\diag}{\text{diag}}

\newcommand{\projOm}[1]{\text{proj}_{\bm \Omega}\left[#1\right]}
\newcommand{\projOmh}[1]{\text{proj}_{\Omega^h}\left[#1\right]}
\newcommand{\col}[1]{\text{col}\left(#1\right)}
\newcommand{\dims}[1]{\text{dim}\left(#1\right)}
\newcommand{\diags}[1]{\diag\left(#1\right)}
\newcommand{\Rset}[1]{\mathbb{R}^{#1}}

\newcommand{\omegaset}[1]{\Omega^{#1}}
\newcommand{\Iset}{\mathcal{I}}
\newcommand{\Hset}{\mathcal{H}}
\newcommand{\Tset}{\mathcal{T}}
\newcommand{\sigmav}{\sigma_{V}}

\newcommand{\aoo}{\sqspecAalpha}

\newcommand{\vi}[1]{\text{VI}(#1)}

\newcommand{\xhstar}{x^{h,*}}

\newcommand{\nhg}{n_h^{\text{g}}}
\newcommand{\nhs}{n_h^{\text{s}}}
\newcommand{\Atau}[1]{A_{\tau}(#1)}

\newcommand{\NE}{^{\text{NE}}}
\newcommand{\Isethg}{\Iset_h^{\text{g}}}
\newcommand{\Iseths}{\Iset_h^{\text{s}}}

\newcommand{\sqspecAalpha}{\sqrt{\rho(A(\alpha))}}
\DeclareMathOperator*{\argmin}{arg\,min}

\begin{document}

\title{Projected gradient-tracking in multi-cluster games and its application to power management

\author{Jan Zimmermann, Tatiana Tatarenko, Volker Willert and J\"urgen Adamy}
        % <-this % stops a space
\thanks{}}% <-this % stops a space}
%\thanks{Manuscript received April 19, 2021; revised August 16, 2021.}}



\maketitle

\begin{abstract}
We are concerned with a distributed approach to solve multi-cluster games arising in multi-agent systems. In such games, agents are separated into distinct clusters. The agents belonging to the same cluster cooperate with each other to achieve a common cluster goal while a non-cooperative game is played between the clusters. To be able to deal with the sparsity of information, as each agent only knows a specific part of the problem, we combine gradient-tracking and consensus methods for information distribution into an algorithm that can solve both the cooperative and non-cooperative problem in a single run. The constraints of the problem are taken into account by the corresponding projection operators and linear convergence is proven given an appropriate constant step size. The algorithm is applied to a day-ahead power management problem, posed as a multi-cluster game, and its efficiency is demonstrated by simulations.
\end{abstract}


\section{Introduction}
\IEEEPARstart{D}{istributed} optimization in cooperative and non-cooperative multi-agent systems has been intensively studied over last years \cite{Bianchi2020}, \cite{Bianchi2021}, \cite{Pavel2020}, \cite{Zimmermann2020}, \cite{Tatarenko2019}. A major  
field of application is energy management in Smart Grids \cite{Chen2018, Zhao17}. Leveraging distributed optimization methods in this field is in coherence with the ongoing decentralization trend of energy and power networks. In this context, distributed optimization methods make it possible to achieve the same optimal results as traditional, centralized optimization procedures, without requiring a central computing node that has access to all information about the problem. With this, information privacy of optimization participants is guaranteed. This is of importance in cooperative optimization as it avoids revelation of all information at some single point of intrusion. However, such privacy restrictions are even more essential in non-cooperative scenarios where local information needs to be kept from competitors. 
%Furthermore, distributed methods directly incorporate the communication between the optimization agents into the optimization procedure.\\
In the context of Smart Grids, cooperative optimization has been intensively employed in the context of economic dispatch \cite{Zimmermann2020}, \cite{Tatarenko2019}, \cite{Chen2018} and optimal power flow problems \cite{Wang2017}, while energy management and energy market situations  are often modelled by non-cooperative game theory \cite{Belgioioso2020},\cite{Gabriel2013}. However, less research has been dedicated to scenarios, in which both cooperative and non-cooperative situations arise. Consider a network that consists of a pool of microgrids. Each microgrid encompasses several decentralized generators, both renewable energy plants and fuel based ones, storage devices and consumers. Suppose each microgrid is operated by a different company that makes investments into the microgrid in order to generate profit while supplying the consumers demand. As all plants within a microgrid belong to the same institution, the problem of finding an optimal dispatch for generators and storage is to be solved cooperatively such that the revenue of the microgrid owner can be maximized. Additionally, all microgrids are connected to a main grid from which they can purchase energy to meet demands. This main grid is operated by a different institution, which sets the power price depending on the overall demand of the microgrids. Therefore, the microgrids compete against each other in a power market, run by the main grid operator. Such market situations are modelled by non-cooperative games. Thereby, inside each microgrid, the cooperative problem of dispatch and supply is to be solved, which is, however, dependent on the outcome of the non-cooperative market problem between the microgrids. In order to avoid iterating back and forth between solutions of the cooperative and non-cooperative problems above, a distributed algorithm needs to be employed that can deal with both problems at the same time scale.\\
The above scenario can be modelled by a multi-cluster game, also called $N$-coalition (cluster) game. In such games, the agents of some multi-agent system are allocated into different groups, the so called clusters. The agents inside every cluster collaborate with each other to achieve some common goal, in the  example above, this goal corresponds to the power provision inside the microgrids, while the groups of agents in form of a cluster compete against other clusters, e.g. regarding the commonly used source such as the power from the main grid.\\ %\vspace*{-\baselineskip}
%\subsection{Related work}
There already exists a body of research concerning continuous-time solution algorithms both for finding Nash equilibria \cite{Ye2020} and generalized Nash equilibria \cite{Sun2021} in multi-cluster games. However, less research has been dedicated to discrete-time algorithms, which do not need to be discretized before calculations are run on a digital device. Previously presented methods in the discrete-time domain can be further subdivided into gradient-based and gradient-free algorithms. The latter optimization methods are required when no functional form of the cost functions are available to agents. Such scenarios arise, for example, in game-theoretic energy market problems, where the market operator determines a price based on the bids of market participants. To deal with such settings, the work \cite{Tatarenko2021}, for example, employs a single point gradient-estimation technique and provides an efficient procedure converging to a solution in strictly monotone multi-cluster games.  
However, due to the limitation caused by zero-order information (cost functions' values at a current joint action), gradient-free methods exhibit slow convergence rates.\\
To address fast convergence to a solution, in this paper, we present a gradient-based algorithm. This procedure relies on the gradient-tracking technique, which is originally developed for cooperative distributed optimization problems \cite{Qu2018}, \cite{Pu2021}. The gradient-tracking update procedure allows for a constant step size and for most existing work, discussed below, the convergence rate is linear, given strongly convex objectives. Additionally to these advantages, gradient-tracking can intuitively be adapted from cooperative, distributed optimization problems to the multi-cluster case by restricting the tracking of the gradient to agents that are within the same cluster. However, employing techniques for considering constraints in the optimization process, such as projection methods, is not straightforward. In fact, to the best of our knowledge, so far, there exists a single work \cite{Dong2020} that combines the projection methods with a gradient-tracking approach for distributed optimization by integrating a second step size to weight the projected update by the estimated gradient with the consensus update. In the work \cite{Falsone2022}, the authors combine the gradient-tracking technique with a proximal minimization method for solving cooperative optimization problems. While the resulting algorithm can deal with locally constrained cost functions, a local optimization problem needs to be solved by each agent at every iteration, which leads to an increase of computational complexity for most applications. The convergence rate ana\-lysis of the proximal-tracking method presented in \cite{Falsone2022} and its applicability to multi-cluster games remain open problems.\\
In contrast to the discussed gradient-free methods, the subsequent gradient-based algorithms require inter-cluster communication as it is necessary for the agents to estimate the joint strategy in order to evaluate their local gradients at that point. Therefore, in \cite{Meng2020}, a leader-follower hierarchy is introduced in the clusters, such that inter-cluster communication is restricted to communication between cluster leaders, while the followers within each cluster communicate only with their in-cluster neighbors and with their leader. All communication channels are considered to be undirected. In \cite{Zimmermann2021}, the results of \cite{Meng2020} are generalized for directed communication architectures without the limitation to a leader-based information exchange between the clusters, which enables a faster convergence. Similar algorithms are proposed in \cite{Pang2020} and \cite{Zhou2021} for unconstrained multi-cluster games with directed communication architectures. Thus, there already exist several versions of gradient-based algorithms for solving multi-cluster games in a distributed manner. However, to the best of our knowledge, non of them takes into account potentially existing constraints. However, the ability to deal with constraints is a crucial property of a method to be applicable to multi-cluster games arising in many applications including power management in Smart Grids described above.\\ 
%\vspace*{-\baselineskip}
% In \cite{Meng2020}, a leader-follower hierarchy inside the clusters is introduced and the solution of an unconstrained Multi-Cluster-Game is obtained by. 
%
%
%
%In \cite{Zimmermann2020}, the results of \cite{Meng2020} were generalized for directed communication architectures without the limitation of leader communication between the clusters.
%All works above that employ a gradient-tracking scheme are designed for unconstrained optimization or game-theoretic problems. In fact, there only exist few prior works that apply gradient-tracking procedures in constrained optimization. In \cite{Dong2020}, a projection is used to ensure that the update steps of an optimization procedure for cooperative, distributed optimization lie within a constrained set that is known by every agent. In the respective algorithm, two step-size are employed, one that weights consensus and projected gradient-step updates against each other and another that weights the local gradient-steps in the gradient-tracking update. Convergence is proven for communication over undirected graphs. 
%\subsection{Contributions}
The contributions of the work at hand are as follows:
\begin{enumerate}
	\item We provide a gradient-based method for finding Nash equilibria in a class of multi-cluster games with constrained action sets. This augments the related methods of \cite{Meng2020}, \cite{Zimmermann2021} to constrained optimization problems. 
	\item As multi-cluster games are a generalization of distributed cooperative optimization problems (where all agents are contained within a single cluster), this paper extends the existing literature on cooperative optimization approach as well. In contrast to the work of \cite{Dong2020}, requiring two step sizes for the proposed projected gradient-based method, we provide the convergence proof for our procedure for which a single positive step size with less restrictive bounds needs to be set up. Compared to the work \cite{Falsone2022}, where an extra routine of estimating proximal point iterations needs to be done, our algorithm requires less computational effort of the agents at each time step. Moreover, we provide an estimation for the convergence rate of the proposed algorithm. 
	\item By incorporating constraints into the optimization procedure, we enable applicability of gradient-tracking based methods to multi-cluster games in power management. In this work, we provide simulation results on a specific scenario of this problem type.
\end{enumerate}
%In this work, we extend previous results on gradient-tracking procedures for solving unconstrained Multi-Cluster Games such that constraints imposed by the clusters can be accounted for. This is a crucial step for applying these methods
%To the best of our knowledge, this is the first existing discrete-time, gradient-based algorithm for finding Nash equilibria in Multi-Cluster-Games. 
%Gradient-Tracking with projection both for Multi-Cluster games and distributed optimization. Multi-cluster games are a generalization of Game Theoretic problems and distributed optimization problems. 
%While \cite{Dong2020} needs two different step-sizes $\alpha, \beta$ that need to be chosen in accordance to each other, while we recover the same structure with a single step-size  as in classical, unconstrained gradient-tracking procedures. 
%\vspace*{-\baselineskip}
%\subsection{Structure of the paper}

The paper is structured as follows: In Section~\ref{sec:notation_and_graphs}, we  briefly introduce relevant notations. In Section \ref{sec:formulation_of_problem}, we provide a detailed problem formulation.  We present our algorithm in Section~\ref{sec:algorithm}. Convergence of the algorithm is rigorously proven in the subsequent Section~\ref{sec:convergence}. In Section~\ref{sec:simulations}, we present the power management problem  mentioned in the introduction and provide simulation results for the proposed procedure.  Section~\ref{sec:conclusion} concludes the paper.  

\section{Notations}\label{sec:notation_and_graphs}
Let  $\nabla_h F(x^h, y)$ denote the gradient $\nabla_{x} F(x, y) |_{x = x^h}$, where $h$ is some identifier and not an exponent. 
We use bold face to differentiate between vectors $x \in \Rset{n}$ and stacked vectors $\x \in \Rset{Nn}$, such that $\x = \col{x_1, x_2, ..., x_N}$, where $x_i \in \Rset{n}$. Let $\1_N$ denote a vector of dimension $N$, where all entries are equal to 1. $||\cdot||$ denotes the Euclidean norm. The operator $\diags{A_1,A_2, ..., A_n}$ creates a block diagonal matrix with block elements $A_1, ... , A_n$. Let $(x_i)_{i \in \Iset} = (x_i)_{i=1}^n$ with ordered index set $\Iset \in \lbrace 1, 2, ..., n \rbrace$  be an ordered sequence of elements $x_1, x_2, ...,x_n$. The symbol $\otimes$ denotes the Kronecker product. The projection of vector $u$ onto set $\Omega$, $\text{proj}_\Omega \left[u\right] = \argmin_{\omega \in \Omega} ||\omega - u|| $, provides the vector $\omega$ inside $\Omega$, which has the smallest Euclidean distance to $u$.  \\
We consider undirected graphs $\mathcal{G}(\Vset, \Eset)$ with vertex set $\Vset$ and edge set $\Eset = \Vset \times \Vset$. Each vertex $i \in \Vset$ represents an agent of a multi-agent system, while the communication channels between two agents $i$ and $j$ are modelled as edges $(i,j) \in \Eset$. Note that $(i,j) \in \Eset \iff (j,i) \in \Eset$, because the graph edges are undirected. As a result, the associated adjacency matrix $A \in \Rset{|V| \times |V|}$ of graph $\mathcal{G}$ is symmetric. A graph is considered to be connected if there exists an undirected path from vertex $i$ to  $j$ for all pairs $i, j \in \Vset$. \\
Let $\mathcal{X} \subseteq \Rset{n}$. The mapping $M:\mathcal{X} \rightarrow \Rset{n}$ is strongly monotone with constant $\mu>0$, if $(x - y)^T(M(x) - M(y)) \geq \mu || x - y ||^2$ for all $x,y \in \mathcal{X}$. The function $f: \mathcal{X} \rightarrow \Rset{}$ is Lipschitz-continuous with constant $L>0$, if $||f(x) - f(y)|| \leq L ||x - y||$ for all $x,y \in \mathcal{X}$. \\
Given a mapping $F(\cdot):\Rset{n}\to\Rset{n}$, the fixed point of $F$ is such point $x\in\Rset{n}$ that $x=F(x)$.

\section{Problem Formulation}\label{sec:formulation_of_problem}
In this section, we first formulate the multi-cluster game-theoretic problem  and then introduce the information setting under which this problem needs to be solved. Note that we differentiate in the following the number of agents by capital letter $N$ form the number of dimensions $n$. \vspace*{-0.4cm}
\subsection{Game-theoretic Problem}
Let us consider a network consisting of $N$ agents, indexed by $1,\ldots, N$ and contained in set $\mathcal{I} = \lbrace 1, ..., N\rbrace$, which are separated into $H$ clusters, contained in the set $\mathcal{H} = \lbrace 1, ..., H \rbrace$,  such that each cluster $h$ contains $N_h$ agents. The set $\mathcal{I}_h$ contains all agents that are part of cluster $h$. It is assumed that no agent is part of more than one cluster, i.e. $\Iset_h \cap \Iset_l = \emptyset, \ \forall  h,l \in \Hset, h \neq l,$ and each cluster contains at least one agent, i.e. $\Iset_h \neq \emptyset$.  The agents within a cluster cooperate with each other to achieve the cluster's goal, while the clusters compete against each other in a non-cooperative game. In this game, the clusters are virtual players, as the actual decisions in favour of the respective cluster goal are determined by the agents that are associated with the cluster. In consequence, the agents of cluster $h$ are only allowed to change the strategy vector of their own cluster, defined by vector $x^h \in \mathbb{R}^{n_h}$. 
The objective of each cluster $h$ is mathematically expressed by the minimization of function $F^h(x^h, x^{-h})$, which depends both on the cluster's strategy $x^h$ and the combined strategies of all other clusters $x^{-h}$. In such a multi-cluster game, the cost function $F^h$ is equal to the sum of local cost functions of agents belonging to cluster $h$, i.e. $F^h(x^h, x^{-h}) = \sum_{i = 1}^{n_h}  f_i^h(x^h, x^{-h})$, where $f_i^h: \Rset{n} \to \Rset{}$ with $n = \sum_{h=1}^Hn_h$. Additionally, constraints are considered, which restrict the search space of the optimization process such that the strategy $x^h$ of each cluster $h$ needs to lie within some cluster's constraint set $\Omega^h$. Regarding this set, we make the following assumption.
\begin{assum}\label{as:setX}
	Each constraint set $\omegaset{h}$, $ \forall h \in \Hset$, is convex and compact. 
\end{assum}
The overall strategy vector of the game is denoted by vector $x = \col{x^h, x^{-h}} \in \Rset{n}$.  With these definitions at hand, the optimization problem of each cluster $h$ is formulated as follows:
%\begin{subequations}
	\begin{align}\label{prob:mcg}
		&\min_{x^h} F^h(x^h, x^{-h}) = \min_{x^h} \frac{1}{N_h}\sum_{i=1}^{N_h} f_i^h(x^h, x^{-h}), \cr
		&\text{s.t. }\,  x^h \in \omegaset{h}.
	\end{align}
%\end{subequations}
To further specify the problem class under consideration, we make the following assumption.
\begin{assum}\label{as:local_costfunction}
	All local cost functions $f_i^h(x^h, x^{-h}),$ $\forall i \in \Iset_h, \ \forall h \in \Hset,$ are defined on the whole $\Rset{n}$, continuously differentiable and convex in $x^h$ for any fixed $x^{-h}$. 
\end{assum}
\begin{remark}\label{rem:lipschitz_continuity}
	In combination, Assumptions~\ref{as:setX} and \ref{as:local_costfunction} imply Lipschitz continuity  of the local cost functions over the corresponding set $\omegaset{h}$ with some constant $L_1$ as their gradients are bounded on this compact set $\omegaset{h}$. Both assumptions are standard in the relevant literature on distributed Nash equilibrium problems \cite{Bianchi2020}, \cite{Bianchi2021}, \cite{Pavel2020}, \cite{Tatarenko2021}.
\end{remark}
Let the Cartesian product of all constraint sets be denoted by $\Omega = \omegaset{1} \times ... \times \omegaset{H}$.  
In order to solve not only the cooperative optimization problem, but also to find a stable strategy between the clusters, the non-cooperative game $\Gamma(H, \Omega, \lbrace F^h \rbrace)$ between the clusters needs to be analyzed. An important component of any game satisfying Assumption~\ref{as:local_costfunction} is the so called game mapping $M:\Rset{n} \rightarrow \Rset{n}$  defined as follows:
\begin{equation}\label{eq:mapping}
		M(x) = \col{(\nabla_h F^h(x))_{h \in \Hset})},
\end{equation}
where $\nabla_h F^h(x) = \frac{1}{N_h} \nabla_h \sum_{i=1}^{n_h} f_i^h(x^h, x^{-h})$. In this work, we focus on games, whose mappings are strongly monotone.
\begin{assum}\label{as:mapping}
	The game mapping $M(x)$ is strongly monotone over $\Rset{n}$  with some constant $\mu>0$ and Lipschitz continuous  over $\Rset{n}$ with some constant $L_0 > 0$.  
\end{assum}
Note that the assumption above is a standard assumption in the literature dealing with fast converging methods for Nash equilibria seeking~\cite{TatShiNedich21}. 

We consider a \emph{Nash equilibrium} in game $\Gamma (H, \Omega, \lbrace F^h \rbrace)$  as a stable outcome in the multi-cluster game under consideration and, thus, providing a solution to the optimization problem in~\eqref{prob:mcg}.
\begin{definition}\label{def:NE}
	A point  $x^{\text{NE}}\in\Omega$ is called a \emph{Nash equilibrium} of the game $\Gamma(H, \Omega, \lbrace F^h \rbrace)$ if $F^h((x^h)\NE, (x^{-h})\NE) \leq F^h(x^h, (x^{-h})\NE)$, for all $x^h \in \omegaset{h}$, $\forall h \in \Hset$.
\end{definition}
Our goal is to learn such a stable action in a game through designing an appropriate algorithm taking into account the information setting in the system. 
%To do so, we first connect  Nash equilibria for $\Gamma (H, \Omega, \lbrace F^h \rbrace)$ with solutions of a corresponding variational inequality.
%\begin{definition}\label{def:vi}
%	The solution set to the variational inequality $\vi{\Omega, M}$, associated with the game  $\Gamma(H, \Omega, \lbrace F^h \rbrace)$ is defined by the set $SOL(\Omega, M) = \{x^*\in\Omega: \, (x - x^*)^T M(x^*) \geq 0, \ \forall x \in \Omega \}$.
%\end{definition}
%We can formulate the following connection between the Nash equilibria and the set $SOL(\Omega, M)$.
%\begin{theorem}\label{th:VINE}(\cite[Proposition~1.4.2]{FaccPang1})
%	Let Assumptions~\ref{as:setX} and~\ref{as:local_costfunction} hold in game $\Gamma(H, \Omega, \lbrace F^h \rbrace)$ . Then, some vector $x^*\in \Omega$ is a Nash equilibrium in $\Gamma$, if and only if $x^*\in SOL(\Omega, M)$. Moreover, the Nash equilibrium $x^*$ exists. If Assumptions~\ref{as:mapping} holds, the Nash equilibrium in $\Gamma$ exists and is unique.
%\end{theorem}
\vspace*{-0.4cm}
\subsection{Distributed setting}
In this section, we present the information setting in the system and provide a reformulation of problem~\eqref{prob:mcg} which takes into account this setting. First, we declare all local cost functions  to be private such that only agent $i$ of cluster $h$ knows the function $f_i^h(x^h, x^{-h})$. Additionally, the constraint sets $\Omega^h$, $h \in \Hset$ are only known by the agents of the respective cluster $h$. As a consequence of these privacy restrictions, the agents within each cluster need to coordinate themselves to achieve the true minimum of the cluster function. Therefore, a communication architecture is established between the agents of the same cluster, which is represented by a communication graph. We denote the graph connecting the agents in cluster $h$ by $\Gset^h(\Vset^h, \Eset^h)$ and make the following assumption.
\begin{assum}\label{as:local_graphs}
	The local graphs $\mathcal{G}^h$, connecting the agents inside cluster $h$, are undirected and connected. The associated weight matrices $V^h$ are double stochastic, i.e. $\1^T V^h = \1^T$ and $V^h \1 = \1$, and have positive diagonal elements $v^h_{ii} > 0$ $ \forall i \in \Iset_h$. 
\end{assum}
Additionally, a global communication architecture is establish, which connects agents independently of their cluster membership. This architecture is described by graph $\Gset(\Vset, \Eset)$, where $\Vset = \Iset$. Similar to the local graphs, we make the following assumption.
\begin{assum} \label{as:global_graph}
	The global graph $\mathcal{G}$, connecting all agents regardless of cluster association, is undirected and connected. The associated weight matrix $W$ is double stochastic, i.e. $\1^T W = \1^T$ and $W \1 = \1$, and has positive diagonal elements $w_{ii} > 0$ $\forall i \in \Iset$. 
\end{assum}
In order to evaluate the gradient of their local cost functions, each agent $i$ maintains an estimation $x_i  \in \Rset{n}$ of the joint strategy vector $x \in \Rset{n}$. The communication architecture allows the agents to exchange information about their current strategy estimations with other agents in the system. Each agent's estimation is comprised of its estimation regarding the cluster strategy $x_i^{(h)} \in \Rset{n_h}$ and of the estimations regarding the strategies of all other clusters $x_i^{(-h)} \in \Rset{n - n_h}$, such that $x_i = \col{x_i^{(1)}, ..., x_i^{(h)},..., x_i^{(H)}}$. The estimation of all agents are combined in the vector $\x = \col{(x_i)_{i \in \Iset}} \in \Rset{Nn}$.\\
Likewise, we define vector $\x^h$, which contains the estimations made by agents that belong to cluster $h$, i.e. $\x^h = \col{(x_i)_{i \in \Iset_h}} \in \Rset{N_hn}$, and the vector $\x^{h(h)} = \col{(x_i^{(h)})_{i \in \Iset_h}} \in \Rset{N_h n_h} $. The vector $\x^{h(-h)}$ is defined accordingly. Note that in the previous definitions we differentiate between estimations regarding cluster $h$ by superscript $(h)$ and membership of cluster $h$ by superscript $h$. \\
Therefore, induced by the privacy of information, the problem in \eqref{prob:mcg}  takes the form \vspace*{-0.3cm}
	\begin{align}\label{prob:mcg_extended}
		&\min_{\x^{h(h)}}  \F^h(\x^{h(h)}, \x^{h(-h)}) = \min_{\x^{h(h)}} \frac{1}{N_h}\sum_{i=1}^{N_h} f_i^h \left(x_i\esth, x_i\estmh\right),\nonumber\\
		&\text{s.t. } \, \x^{h(h)}  \in \bm \Omega^h, (\I_{Nn} - \W)\x = 0.
	\end{align}
with $\bm \Omega^h := \lbrace \x^{h(h)} | x_i\esth \in \Omega^h, \ \forall i \in \Iset_h \rbrace $.
The consensus condition $(\I_{Nn} - \W)\x = 0$ with  $\W = W \otimes \I_n$ and some double-stochastic matrix $W \in \Rset{N \times N}$ ensures that $x_i = x_j, \ \forall i,j \in \Iset$, i.e. the agents reach a consensus regarding their estimations. 
As a result of the additional estimation problem, the dimension of the optimization problem increases by the factor $N$. Accordingly, the mapping of the multi-cluster game can be extended to the  mapping $\M: \Rset{Nn} \rightarrow \Rset{ \sum_{h=1}^H N_h n_h}$ such that
\begin{equation}\label{eq:exmapping}
	\M(\x) = \col{(\1_{N_h} \otimes \nabla_{h} \F^h(\x^h) )_{h \in \Hset}},
\end{equation}
with $\nabla_{h} \F^h(\x^h) = \frac{1}{N_h}  \sum_{i = 1}^{N_h}\nabla_{h}  f_i^h \left(x_i\esth, x_i\estmh\right)$.  
Note that this mapping is no longer strongly monotone compared to its original counterpart $M(x)$ in~\eqref{eq:mapping} (see Assumption~\ref{as:mapping}). To ease the notations, we define the assignment matrix $\Q$, using $n_{<h} := \sum_{j=1}^{h-1} n_j$ and $n_{>h}:=\sum_{j=h+1}^H n_j$, such that
\begin{align}
	&\Q_i^h =  [0_{n_h \times n_{<h}}, \I_{n_h}, 0_{n_h\times n_{>h}}], \nonumber \\
	&\Q^h =  \diags{(\Q_i^h)_{i \in \Iset_h}}  \text{ and }	\Q =  \diags{(\Q^h)_{h \in \Hset} }. \label{eq:Q}
\end{align}
%with dimensions $\dims{\Q_i^h} = n_h \times n$, $\dims{\Q^h} = N_h n_h \times n N_h$ and $\dims{\Q} = \sum_{h=1}^{H} N_h n_h \times nN$. Note that
Note that $\Q_i^h x_i = x_i\esth$, $\Q^h \col{(x_i)_{i\in I_{h}}} = \x^{h(h)}$  and $\Q \x =  \col{(\x^{h(h)})_{h \in \Hset}}$. Therefore, by multiplying the estimation vector $\x$ by $\Q$, each agent's estimation regarding their own cluster's variable is extracted in a stacked vector.\\
To  clarify the connections of the original multi-cluster problem and its extension by estimation described in  \eqref{prob:mcg_extended}, we state some equivalences between the solutions of both problems and their connections to the specific variational inequality. 
\begin{lemma}\label{lemma:equivalences}
	Let Assumptions~\ref{as:setX}, \ref{as:local_costfunction} and \ref{as:mapping} hold. Then, there exists a unique Nash equilibrium of  the game $\Gamma(H, \Omega, \lbrace F^h \rbrace)$. Moreover, the following statements are equivalent:\vspace{-0.1cm}
	\begin{enumerate}[label=\alph*)]
		\item $x^*$ is the unique solution of the variational inequality $\vi{\Omega, M}$, i.e. $(x - x^*)^T M(x^*) \geq 0, \ \forall x \in \Omega$.
		\item $x^*$ is a unique Nash equilibrium of  the game $\Gamma(H, \Omega, \lbrace F^h \rbrace)$.
		\item $\xstar= \1_N \otimes x^*$ is a fixed point of the mapping $\Hb(\x) = \projOm{\x - \alpha \Qt \M(\x)}$. 
		\item $\xstar= \1_N \otimes x^*$ is a stable solution of the extended multi-cluster game \eqref{prob:mcg_extended}.
	\end{enumerate}
\end{lemma}\vspace{-0.2cm}
Proof can be found in Appendix \ref{ap:proof_equivalences}. \\
Before we present our algorithm for solving the extended multi-cluster game, we state well-known, nevertheless important norm inequalities regarding the weight matrices $W$ and $V^h$. 
Let Assumptions~\ref{as:local_graphs} and \ref{as:global_graph} hold respectively, then there exists a $\sigma \in (0,1)$ and $\sigma_V = \max\lbrace \col{(\sigma_V^h)_{h \in \Hset}}\rbrace \in (0,1)$ such that for any $x \in \Rset{N}$ and $y \in \Rset{N_h}$
\begin{align}
\nl W x -  \1_N  \bar{x} \nr &\leq \sigma \nl  x -  \1_N  \bar{x} \nr \label{eq:sigmaw}, \\
\nl V^h y - \1_{N_h} \bar{y} \nr &\leq \sigmav \nl  y  - \1_{N_h} \bar{y} \nr, \label{eq:sigmav}
\end{align}
where $\bar x = \frac{1}{N} \1^T_N x$ and $\bar{y} =  \frac{1}{N_h} \1^T_{N_h} y$.

\section{Algorithm}\label{sec:algorithm}
 Classical agent based approaches to finding Nash equilibria in some convex game $\Gamma (H,  \Omega, \lbrace F^h\rbrace )$ employ projected gradient schemes, where each player $h$ updates its action according to the following procedure:
\begin{equation}\label{eq:central}
	x^h(k+1) = \text{proj}_{\Omega^h}\left[{x^h(k) - \gamma \nabla_h F^h(x)} \right].
\end{equation} 
Under Assumptions~\ref{as:setX}-\ref{as:mapping} and given an appropriate choice of the step size $\gamma>0$, the iterates above converges to a Nash equilibrium in $\Gamma (H,  \Omega, \lbrace F^h\rbrace )$ with a linear rate (see, for example, \cite{Facchinei2007}). However, for multi-cluster games this scheme is not applicable as it is, because the agents belonging to cluster $h$ are not able to evaluate the cluster gradient $\nabla_h F^h(x)$ due to their information restrictions. Additionally, no agent has direct access to the decisions of all other agents in the system. Therefore, the standard projected gradient scheme needs to be augmented by two estimation processes: First, by a consensus process that estimates the actions of other cluster but also leads to an agreement among agents of the same cluster. Secondly, by an estimation process regarding the cluster gradient $\nabla_h F^h$. For the latter, we leverage the gradient-tracking technique (see for example \cite{Qu2018}) to estimate the cluster gradient in the auxiliary variable $y_i^h(k) \in \Rset{n_h}$, which is maintained by every agent $i$ belonging to cluster $h$ and initialized with $y_i^h(0) = \nabla_h f_i^h(x_i(0))$. 
We propose the following updates for each agent $i$, $k = 0, 1, ...$:%\vspace*{-\baselineskip}
\begin{subequations}\label{alg:agentwise}
	\begin{align}
	&\hat{x}_i(k) = \sum_{j=1}^{N} w_{ij} x_j(k),  \label{alg:agentwise_consensus} \\
	&x_i^{(h)}(k+1) = \projOmh{\hat{x}_i^{(h)}(k) - \alpha y_i^h(k)} , \label{alg:agentwise_gradientstep} \\
	&x_i^{(-h)}(k+1) = \hat{x}_i^{(-h)}(k),  \label{alg:agentwise_estotherclusters} \\
	&y_i^h(k+1) = \sum_{j=1}^{N_h} v^h_{ij} y_i^h(k)\! + \! \nabla_h f_i^h(x_i(k+1)) \!-\! \nabla_h f_i^h(x_i(k)), \label{alg:agentwise_gradienttracking}
	\end{align}
\end{subequations}
where $\alpha > 0$ is some step size. In Equation~\eqref{alg:agentwise_consensus}, agent $i$ updates its estimation by a weighted sum of its own estimation and the estimations of its direct neighbors, which send their estimations to $i$ over the global communication graph $\mathcal{G}$. The weights correspond to the entries of weight matrix $W$, associated with graph $\mathcal{G}$. While the estimations $\hat x^{(-h)}$ regarding other clusters are not altered, because agent $i$ has no influence on the strategy choice of other clusters (see \eqref{alg:agentwise_estotherclusters}), in Equation ~\eqref{alg:agentwise_gradientstep} a gradient step is performed in the direction of the estimated cluster gradient $y_i^h(k)$. In Equation~\eqref{alg:agentwise_gradienttracking} the cluster gradient is estimated. Here, the agents first calculate a weighted sum of their own estimations and the estimations of their direct neighbors in cluster $h$ using the local communication graph $\mathcal{G}^h$ and the associated weight matrix $V^h$. Then, the gradient of the local cost functions is evaluated at the strategy estimations $x_i(k+1)$ and $x_i(k)$ and the difference is added to the results of the inner-cluster consensus step.\\
By defining the gradient-tracking vectors $\y^h = \col{(y_i^h)_{i\in \Iset_h}},\ \y  = \col{(\y^h)_{h \in \Hset}} $ 
and the vectors of local gradients $\G^h(k) = \col{(\nabla_h f_i^h(x_i(k))_{i \in \Iset_h})}$ and $\G(k) = \col{(\G^h(k))_{h \in \Hset}}$, we are able to summarize the update equations of all agents of the system in the algorithm 
\begin{subequations}\label{alg:vectorwise}
	\begin{align}
		\xhat(k) &= \W \x(k), \label{alg:vectorwise_consensus} \\
		\x(k+1) &= \projOm{\xhat(k) - \alpha \Qt \y(k)}, \label{alg:vectorwise_x} \\
		\y(k+1) &= \V \y(k) + \G(k+1) - \G(k),\label{alg:vectorwise_y}
	\end{align}
\end{subequations}
where $\W = W \otimes \I_n$,  $\V = \diags{(V^h \otimes \I_{n_h})_{h \in \Hset}}$, $\bm \Omega = \lbrace \x \in \Rset{Nn} | \Q\x \in \bm \Omega^h \rbrace $  and $\Q$ as in \eqref{eq:Q}. Note that the last equation \eqref{alg:vectorwise_y} can be split into update equations for every cluster, such that $
\y^h(k) = \V^h \y^h(k) + \G^h(k+1) - \G^h(k), \ \forall h \in \Hset$.\\ 
As motivated in the beginning of this section, the presented algorithm can be considered an extension of the projected gradient scheme by gradient and strategy estimation. This is clearly visible in the vector-form algorithm in \eqref{alg:vectorwise}, where in Equation~\eqref{alg:vectorwise_consensus} the strategy's estimations and in Equation~\eqref{alg:vectorwise_y} the gradients' estimations are performed. For the latter, it can be shown that summing over the gradient tracking variables $y_i^h(k)$ of all agents $i$ that belong to cluster $h$, the direction of the cluster gradient at the time $k$ is recovered, provided that the variables are initialized according to $y_i^h(0) = \nabla_h f_i^h(x_i(0))$, i.e. it holds that 
\begin{equation}
	\sum_{i = 1}^{N_h} y_i^h(k) = \sum_{i = 1}^{N_h} \nabla_h f_i^h(x_i(k)).
\end{equation}
This result follows from the stochasticity of the weight matrices $V^h$ and is a well-known fact for the methods based on gradient tracking technique, see \cite{Pu2021}, \cite{Zimmermann2021}.\\
In Equation~\eqref{alg:vectorwise_x}, the standard projected gradient scheme is recovered, as $\y(k)$ tracks the cluster gradients and assignment matrix $\Q$ does not change the direction of the estimated gradients but rather ensures that appropriate coordinates of the extended estimation vector $\x(k)$ are updated.

\section{Proving convergence}\label{sec:convergence}
Before we start with the convergence proof, we provide some preliminary definitions.
Let the gradient-tracking average be defined as
\begin{align*}
	\bary^h(k) &= \1_{N_h}\!\otimes\!\frac{1}{N_h} \sum_{i = 1}^{N_h} y_i^h(k) =  \1_{N_h}\! \otimes\! \frac{1}{N_h} \sum_{i = 1}^{N_h} \nabla_h f_i^h(x_i(k)).
\end{align*}
Thereby, $\bary^h(k)$ is the average over the gradient-tracking variables of cluster $h$ at time step $k$, repeated $N_h$ times into a vector of dimension $\dims{\bary^h} = N_h n_h$.  Let $\bary(k) = \col{(\bary^h(k))_{h \in \Hset}}$ denote the stacked vector. Using the additional assignment matrices
\begin{equation*}
	\bm \R^h = \frac{1}{N_h} \1_{N_h}  \1_{N_h}^T \otimes  \I_{n_h} \text{ and }\ 
	\R = \diags{(\R^h)_{h \in \Hset} }
\end{equation*}
with dimensions $\dims{\R^h} = N_h n_h \times N_h n_h$ and $\dims\R = \sum_{h=1}^H N_h n_h \times \sum_{h=1}^H N_h n_h$, we can write $
	\bary^h(k) = \R^h \y^h(k) \text{ and }	\bary(k) = \R \y(k)$. \\
Several challenges have to be faced when aiming to prove convergence of the algorithm in \eqref{alg:vectorwise}. First, three different dynamics need to be accounted for: The consensus dynamic regarding the estimation of the overall strategy vector and gradient-tracking variables inside the cluster, the estimation of the cluster gradient in the gradient-tracking variables $y_i^h(k)$, and the actual optimization process in form of projected gradient-descent. All these dynamics are interconnected with each other and the associated step size is to be set up  to allow the algorithm to achieve the Nash equilibrium of the multi-cluster problem under consideration. The second challenge concludes in the fact that, in contrast to the original mapping $M(x)$ of the game which is assumed to be strongly monotone (see Assumption~\ref{as:mapping}), the extended mapping $\M(\x)$ in Equation~\eqref{eq:exmapping} does not possess this property. However, it is possible to state the following lemma, which states Lipschitz continuity of $\M(\x)$ and whose proof can be found in \cite{Bianchi2021}.
\begin{lemma} \label{lemma:lipschitz_extended_mapping}
	Under Assumptions~\ref{as:mapping}, the extended mapping $\M(\x)$ is Lipschitz continuous with some constant $L> 0$ such that $\mu \leq L \leq L_0$, where $\mu, L$ are defined in the assumption. 
\end{lemma}

Let now $x^*$ be the Nash equilibrium of the multi-cluster game $\Gamma (H,  \Omega, \lbrace F^h\rbrace )$ and let $\xstar = \1_N \otimes x^*$. Then, according to Lemma~\ref{lemma:equivalences}, $\xstar$ is the solution to the extended multi-cluster game-theoretic problem \eqref{prob:mcg_extended}. Therefore, in order to prove convergence of algorithm \eqref{alg:vectorwise}, we essentially need to show that the distance between the estimations $\x$ and the solution $\xstar$ converges to zero over time, i.e. $\lim_{k \rightarrow \infty }|| \x(k) - \xstar||  = 0$. To do so, we demonstrate that the combined cluster gradient estimation $\y(k)$ converges to the average of the local cost function gradients, i.e. $\lim_{k \rightarrow \infty } || \y(k) - \bary(k)|| = 0$. As it has been pointed out, evolution of the estimations $\x(k)$ depends on that of the variable $\y(k)$ and vice versa. In the following proposition, we show that under the algorithm in \eqref{alg:vectorwise}, $\x(k)$ and $\y(k)$ evolve according to a specific linear inequality system.
\begin{prop} \label{prop:linear_inequality}
	Let Assumptions~\ref{as:setX}, \ref{as:local_costfunction}, \ref{as:mapping}, \ref{as:local_graphs} and  \ref{as:global_graph} hold and let $\xstar = \1_N \otimes x^*$. Then there exists a fixed point of the mapping $\Hb(\x) = \projOm{\x - \alpha \Qt \M(\x)}$ and, based on the algorithm in \eqref{alg:vectorwise}, the  linear inequality system 
	\begin{equation}\label{eq:linineqsys}
		\tau(k+1) \leq A_\tau(\alpha) \tau(k)
		\end{equation} 
		can be established,
with  $\tau(k) = [||\x(k) - \xstar ||, || \y(k) - \bary(k) ||]^T$ and a positive matrix $\Atau{\alpha} \in \mathbb{R}^{2\times 2}$. 
Furthermore,  there exists a positive constant $\overline{\alpha}$, dependent on the parameters $N$, $L$, $L_0$, $\mu$, $\sigma_V$, and $\sigma$, such that  $\rho(\Atau{\alpha}) < 1$ for any  $\alpha \in(0, \overline{\alpha})$.
\end{prop}
The proof is provided in Appendix \ref{ap:proof_linear_inequality}. 
\begin{remark}
	Matrix $\Atau{\alpha}$ and the upper bound $\overline{\alpha}$ are further specified in the proof of Proposition \ref{prop:linear_inequality}. 
\end{remark}
Finally, in the following theorem, we are able to summarize the previous results and rigorously prove convergence of the algorithm in \eqref{alg:vectorwise}. 

\begin{theorem}\label{thm:convergence}
	Let Assumptions~\ref{as:setX}, \ref{as:mapping}, \ref{as:local_graphs} and \ref{as:global_graph}  hold. Then, there exists a unique Nash equilibrium, which is a stable solution of the multi-cluster game problem \eqref{prob:mcg}.\\
	Moreover, let the step size of the algorithm~\eqref{alg:vectorwise} be chosen such that $0 < \alpha < \overline{\alpha}$, where $\overline{\alpha}$ is introduced in Proposition~\ref{prop:linear_inequality}. Then the combined estimation vector converges toward the stacked Nash equilibrium vector $\xstar = \1_N \otimes x^*$, i.e. $\lim_{k \rightarrow \infty } \x(k) = \xstar$ and the convergence is linear with the rate $O(\rho(\Atau{\alpha})^k)$, where $\rho(\Atau{\alpha}) <1$. 
\end{theorem}

\begin{IEEEproof}
	Provided that the mapping $M(x)$ in \eqref{eq:mapping} fulfils Assumption~\ref{as:mapping}, i.e. it is strongly monotone, the Variational Inequality $\vi{\Omega, M}$, associated with the problem in  \eqref{prob:mcg} has a unique solution \cite{Facchinei2007}. According to Lemma~\ref{lemma:equivalences}, the Nash equilibrium of the multi-cluster problem exists in this case and is unique as well.\\
	By Lemma~\ref{lemma:equivalences}, the vector $\xstar$ is a fixed point of the mapping $\Hb(\x) = \projOm{\x - \alpha \Q^T \M(\x)}$. Therefore, Proposition~\ref{prop:linear_inequality} can be invoked, which provides us with the inequality system in Equation \ref{eq:linineqsys}. Furthermore, for the spectral radius it holds that $\rho(\Atau{\alpha}) <1$, if $\alpha$ is chosen in the specified bounds, rendering the autonomous system stable.Therefore, $\lim_{k \rightarrow \infty } \tau(k+1)/\tau(k)< \rho(\Atau{\alpha}) <1$, which proves linear convergence rate $O(\rho(\Atau{\alpha})^k)$ to the desired state $\col{0,0}$. This concludes the proof. 
\end{IEEEproof}
\begin{remark}
	Note that Theorem~\ref{thm:convergence} ensures that all local estimations $x_i(k), \ \forall i \in \Iset$ reach a consensus and that this consensus is a solution to the multi-cluster game.
\end{remark}

\section{Simulations}\label{sec:simulations}
\subsection{Day-ahead power management as multi-cluster game}
As motivated in the Introduction section, we aim to solve a day-ahead power management problem between microgrids, modeled as a multi-cluster game. For the formulation of the game, we loosely rely on \cite{Belgioioso2020}. Let us consider a system consisting of $H$ microgrids. Each microgrid aims to provide power for its consumers with power demand $d^h(t)$ in every time slot $t \in  \lbrace 1, ..., T \rbrace = \mathcal{T}$ of the next day while minimizing the cost of doing so. In order to supply the demanded power,  each microgrid $h$ has two possibilities: Firstly, it can buy power of amount $p^h(t)$ in time slot $t$ from the main grid for the price per power unit
$
	c^{\text{main}}(\bm p(t)) = q\left(\sum_{h=1}^H p^h(t)\right)
$
with $\bm p(t) = \col{p^1(t), ..., p^h(t), ..., p^H(t)}$ and some positive factor $q> 0$. Note that the price is  demand dependent such that the provision strategy from one microgrid has an influence on the cost of all other microgrids. Therefore, if microgrid $h$ buys power of amount $p^h(t)$ in time slot $t$, the costs
\begin{equation}\label{eq:cost_maingrid}
	f_{h}^{\text{mg}}(\bm p(t)) = c^{\text{main}}(\bm p(t)) p^h(t)
\end{equation}
occur. This cost function is the coupling point of the problem, resulting in the non-cooperative game between the microgrids.\\
Secondly, each microgrid can use its own facilities, meaning generators and batteries. Let microgrid $h$ own $\nhg$ generators and $\nhs$ batteries and let $\Isethg$ and $\Iseths$ denote the respective sets. The operating cost of a controllable, fuel-based generator $i$ of microgrid $h$ for producing the amount of power $g_i(t)$ in time slot $t$ is approximated by the quadratic function
\begin{equation}\label{eq:cost_generators}
	f_{i,h}^{\text{g}}(g_i(t)) = a_i^hg_i(t)^2 + b_i^h g_i(t) + c_i^h 
\end{equation}
with constants $a_i^h, b_i^h, c_i^h > 0$. Each generator has lower and upper production capacities, wherefore the combined power production $\gh = \col{(g_i)_{i\in \Isethg}}$, where $g_i = \col{(g_i(t))_{t \in \Tset}}$, needs to lie in constraint set
\begin{equation}\label{eq:constraintset_generators}
	\Omega^h_{\text{g}} := \begin{Bmatrix}
			\gh |& \underline{g_i} \leq g_i(t) \leq \overline{ g_i}, \forall i \in \Isethg, \forall t \in \mathcal{T}  
	\end{Bmatrix}.
\end{equation}
The usage of batteries $j \in \Iseths$ is penalized in every time slot $t$ by the quadratic cost function
\begin{equation}\label{eq:cost_storages}
	f_{j,h}^{\text{s}}(s_j(t)) = a_j^h s_j^2(t) + \text{sgn}(s_j(t)) b_j^h s_j(t) + c_j^h
\end{equation}
with constants $a_j^h, b_j^h, c_j^h > 0$, in order to reduce excessive use of the battery and thus prolong the batteries life time. If $\text{sgn}(s_j(t)) > 0$, then the battery is discharged by the amount $|s_j(t)|$ and if  $\text{sgn}(s_j(t)) < 0$, the battery is charged. The amount of power that can be drawn from or stored in the battery depends on the charge $q_j(t)$ of the battery, which needs to lie within the bounds $0 \leq q_j(t) \leq \overline{q}_j(t)$. We receive the constraint $ -q_j(t) \leq - s_j(t) \leq \overline{q}_j - q_j(t)$.
Each battery looses charge by the leakage effect in between two time slots $t$ and $t+1$ such that $q_j(t+1) = \gamma_j q_j(t)$ with $0 < \gamma_j < 1$ when $s_j(t) = 0$. With this, the charge of battery $j$ at beginning of time slot $t$ can be calculated as $q_j(t) = \gamma_j^{t-1} q_j(1) - \sum_{r= 1}^{t-1} \gamma_j^{t-r}s_j(r)$. Therefore, the constraint $-\gamma_j^{t-1}q_j(1) \leq -  \sum_{r= 1}^{t} \gamma_j^{t-r}s_j(r) \leq \overline{q}_j - \gamma_j^{t-1}q(1)$ needs to hold in every time slot $t$. Additionally, the charge at the end of the planning horizon $q_j(T)$ should be close to $q_j(1)$, i.e. $|q_j(T) - q_j(1)| \leq \epsilon_j$ for some small $\epsilon_j > 0$. Without this constraint, the optimization would lead to depleted batteries at the end of every time period. We summarize these constraints in the constraint set
\begin{align}\label{eq:constraintset_storages}
	\Omega_{\text{s}}^h := \begin{Bmatrix*}[l]
		\multirow{4}{*}{$\bm s^h \Bigg|$}  & \underline{s}_j^h \leq s_j^h(t) \leq  \overline{s}_j^h,   |q_j(T) - q_j(1)| \leq \epsilon_j,  \\
		                                   & -\gamma_j^{t-1}q_j(1) \leq   - \sum_{r= 1}^{t} \gamma_j^{t-r}s_j(r),                                   \\
		                                   & - \sum_{r= 1}^{t} \gamma_j^{t-r}s_j(r) \leq \overline{q}_j - \gamma_j^{t-1}q(1),                        \\
		 &  \forall j \in \Iseths, \forall t \in \mathcal{T},
	\end{Bmatrix*}
\end{align}
with $\bm s^h = \col{(s_j)_{j\in \Iseths}}$ and $s_j = \col{(s_j(t))_{t \in \Tset}}$. 

With the above definition, we can formalize the combined power provision and cost minimization problem of each microgrid $h$ as follows: 
\begin{subequations} \label{prob:microgrid_mcg}
	\begin{align}
		\min_{\bm x^h} &\sum_{t=1}^{T} \left[ f_{h}^{\text{mg}}(\bm p(t))  + \sum_{i = 1}^{\nhg} 	f_{i,h}^{\text{g}}(g_i(t))  +  \sum_{j = 1}^{\nhs}  f_{j,h}^{\text{s}}(s_j(t)) \right] \label{eq:microgrid_mcg_equality_costfunction}\\
		 \text{s.t. } & p^h(t) +  \sum_{i = 1}^{\nhg} g_i(t) + \sum_{j=1}^{\nhs} s_j(t) = d^h(t),  \ \forall t \in  \mathcal{T} \label{eq:microgrid_mcg_equality_constraint} \\
		 &\bm g^h \in \Omega_{\text{g}}^h, \bm s^h \in \Omega_{\text{s}}^h.
	\end{align}
\end{subequations}
with $\bm x^h = \col{\bm p^h, \bm g^h, \bm s^h}$.
For the multi-cluster formulation of above problem, each microgrid is considered a cluster and the agents  are the components of the cluster, i.e. the batteries and the generators within the microgrids. It is assumed that each microgrid is operated by some company. These companies compete against each other regarding the price for power purchased from the main grid, resulting in the non-cooperative game between the distinct grids. Each agent only knows its own cost function, e.g. for some generator $i$, the function $\sum_{t=1}^T f_{h,i}^\text{g}(g_i(t))$ and the cluster constraint set $\Omega^h$ consisting of $\Omega_{\text{g}}^h\cup  \Omega_{\text{s}^h}$ as well as the equality constraint \eqref{eq:microgrid_mcg_equality_constraint}.\\
According to the definition of the cost functions in Equations~\eqref{eq:cost_maingrid}, \eqref{eq:cost_generators} and \eqref{eq:cost_storages}, the resulting cluster cost function is quadratic in each element of the decision variable $\x^h$. Therefore, Assumption~\ref{as:local_costfunction} is fulfilled. It can be checked that the mapping of the game, as defined in Equation~\eqref{eq:mapping} is strongly monotone and therefore fulfills Assumption~\ref{as:mapping}. At last, the constraint set $\Omega^h$ consists of a linear equality constraint and the convex and compact sets $\Omega_{\text{g}}^h$ and $\Omega_{\text{s}}^h$, which therefore satisfies Assumption~\ref{as:setX}. In conclusion, by connecting the agents according to Assumptions~\ref{as:local_graphs} and \ref{as:global_graph}, the algorithm in \eqref{alg:vectorwise} is applicable to the problem defined by~\eqref{prob:microgrid_mcg}.

\vspace*{-\baselineskip}
\subsection{Simulation Setup}
Our simulation setup consists of five microgrids, each containing ten agents, such that the overall agent system is comprised of $N = 50$ agents. Table \ref{tab:sim_setup} shows the distribution of generators and batteries among the microgrids as well as the number of decision variables and constraint functions when considering a time horizon of $T = 24$. In order to show the effects of differing numbers of generators and batteries on the resulting cost, their numbers vary over the microgrids. To make the results for each microgrid more comparable, the demands $d^h(t)$ in all time slots $
t \in \Tset$ inside each microgrid are chosen equally such that $d^i(t) =  d^j(t)$, $\forall i,j \in \Hset$. The demand curve is depicted in Figure \ref{fig:barplot} together with the simulation results and is chosen to resemble the daily consumption of private households. Because of the size of the agent system, we will not go into detail on how we parameterize the cost functions and constraints. However, note that the lower and upper bounds of generators and batteries are chosen such that it is not possible for the microgrids to completely supply themselves with power, i.e. each microgrid is forced to buy power from the main grid in order to be able to satisfy their consumers' demand. This intensifies the competition between the grids. An optimized step size $\alpha = 0.02$ is chosen for each agent by increasing the step-size starting at a small value until the fastest, stable convergence is reached.\\
Global and local communication graphs, $\mathcal{G}$ and $\mathcal{G}^h$ respectively, are chosen such that they fulfill Assumptions~\ref{as:local_graphs} and \ref{as:global_graph}, i.e. all graphs are connected and the weights of the associated matrices $W$ and $V^h$, $\forall h \in \Hset$, were chosen to be double stochastic.\vspace*{-\baselineskip}
\begin{table}[h]
	\centering
	\caption{Overview of microgrid setup and final cost function value $F^h(x^*)$ ($\times 10^8$). MG: microgrid, Gen.: generators, Bat.: batteries, Ag.: Agents, Var.: decision variables, Con.: constraints.}
	\label{tab:sim_setup}
	\begin{tabular}{c|c|c|c|c|c| c}
		        & $\#$ Gen. & $\#$ Bat. & $\#$ Ag. & $\#$ Var. & $\#$ Con. &  $F^h(x^*)$\\ \hline
		 MG 1   &     7     &     3     &    10    &    264    & 627 & 4.36\\
		 MG 2   &     5     &     5     &    10    &    264    & 725 &  6.15 \\
		 MG 3   &     3     &     7     &    10    &    264    & 823 & 7.51 \\
		 MG 4   &     0     &    10     &    10    &    264    & 970 & 9.87 \\
		 MG 5   &    10     &     0     &    10    &    264    & 480 & 3.07\\ \hline
		overall &    25     &    25     &    50    &   1320    & 3625 & 30.96
	\end{tabular}

\end{table}

\begin{figure}[h]
	\centering
	\includegraphics[width=0.8\linewidth]{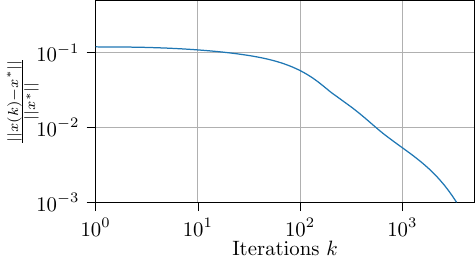}
	\caption{Convergence of the relative error norm between current state  $x(k)$ and Nash equilibrium $x^*$.}
	\label{fig:errornorm}
\end{figure}
\vspace*{-\baselineskip}

\subsection{Results}
\begin{figure*}[h]
	\centering
	\includegraphics[width=0.98\linewidth]{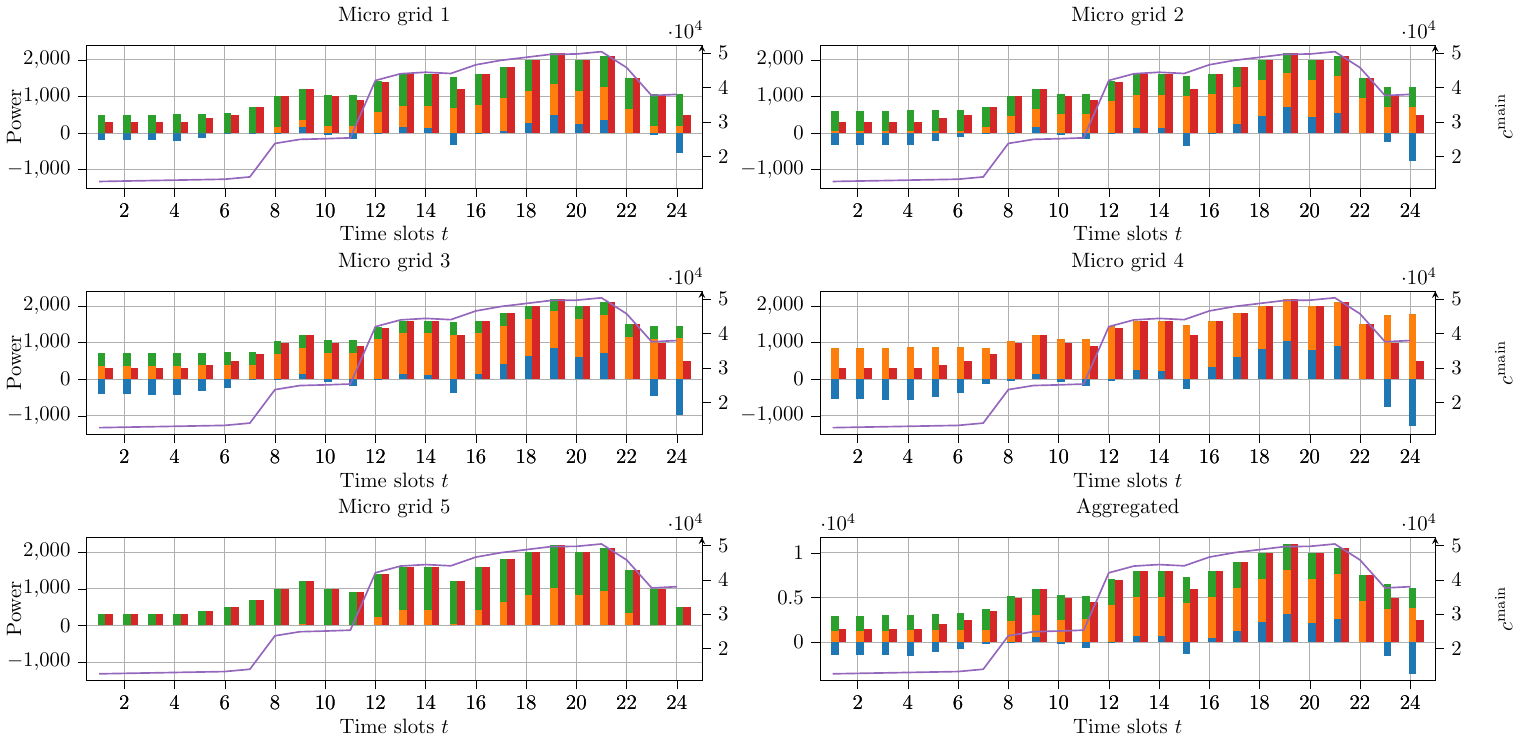}
	\caption{Strategies for {\color{color0} batteries}, {\color{color1} power from main grid}, {\color{color2} power from generators} for microgrids $h = 1, ..., 5$ together with {\color{color3} consumption} in the specific grid. For comparison, the {\color{color4} main grid price function} is plotted as well.}
	\label{fig:barplot}
\end{figure*}

In Figure \ref{fig:errornorm} the convergence of the relative error norm between the current strategy at the first agent of each cluster $x(k) = \col{(x_1^h(k))_{h=1}^H}$ and the true Nash equilibrium $x^*$ is depicted. The Nash equilibrium is calculated by a standard gradient-based algorithm (see~\eqref{eq:central}) under assumption that the full required information is available to the agents. For the specified setup, a relative error of $0.001$ can be achieved after $3392$ iterations. \\
The resulting strategies for microgrids $h = 1, ..., 5$ are depicted in Figure \ref{fig:barplot} together with the aggregated values, i.e. the sum over all strategies in a distinct time slot. Additionally, in each sub figure the main grid price $c^\text{main}$ is plotted. 
Resulting from the choice of generator constraints, each microgrid needs to buy power from the main grid in time slots 17 to 21 where power consumption is high. It can be observed that all microgrids that contain batteries fill their batteries in time slots of low demand, i.e. time slots 1 - 7, as power from the main grid is the cheapest level due to low overall demand.
As expected, the stored power is extracted in times of high consumption,  in order to lessen the dependence on the high main grid prices in these time slots.\\
Interestingly, when examining the resulting costs over all time slots for each microgrid in Table \ref{tab:sim_setup}, increasing the number of generators has a higher impact on the minimization of the cost functions than increasing the number of batteries. For example, microgrid 4 consists entirely of batteries and exhibits the highest cost while microgrid 5 is comprised of generators only and displays the lowest cost. Furthermore, as the number of generators diminishes from microgrid 1 to 3, the costs rise. Therefore, in this scenario, having batteries instead of generators is not beneficial for reducing costs. An explanation for this phenomenon is the high impact of load sharing among a greater number of generators resulting in lower fuel costs. Additionally, when utilizing batteries, the available power in the overall network does not increase; it is only possible to shift power consumption to other time periods. However, if all microgrids with batteries need to charge their batteries in the same time slot, power demand rises and needs to be matched either by the generators or by the main grid. As production of the generators is capped, additional power may need to be drawn from the main grid. This effect can be observed in time slots 23 and 24 right before the end of the planning horizon. Because the charge of each battery may only vary by a small factor $\epsilon = 0.01$ between beginning and end of the time horizon, all batteries need to be recharged after drawing power from them during the high demand period 17 - 21. Although the consumers demand is rather low at the end of the planning horizon, the main grid power price is high due to the power demand in connection with replenishing the batteries. 
Additionally, the batteries leak power over time. In the simulation, $1\%$ of charge was lost in between the time slots, which needs to be resupplied as well. Nevertheless, this does not mean that batteries are useless. The simple fact that the optimal strategy is using batteries means that it is beneficial to have them in the network. However, they need to be used jointly with generators and not instead of the latter.  

%\begin{figure}
%	\centering[h]
%	\includegraphics[width=0.8\linewidth]{img/costfunctions_standalone}
%	\caption{Convergence of the cluster's cost functions.}
%	\label{fig:costfunctions}
%\end{figure}

%\begin{figure}
%	\centering
%	\includegraphics[width=1\linewidth]{img/bar_standalone}
%	\caption{}
%	\label{fig:barstandalone}
%\end{figure}

%
%\left\lbrace \bm s^h \Bigg| \begin{matrix}
%\underline{\bm s}^h \leq \bm s^h \leq \overline{\bm s}^h, \\
%- \bm b^h \leq \bm B^h \bm s^h \leq  \overline{ \bm q}^h -  \bm b^h \\
%- \bm \epsilon + (1 - \bm a^T) \leq \bm a^T \bm s^h \leq \epsilon
%\end{matrix}  \right\rbrace,
%\end{align}

%
% $q_j = \col{(q_j(t))_{t=1}^T}$ and $\bm q^h = \col{(q_j)_{j = 1}^{\nhs}}$ in
%\begin{align}
%	\Omega_{\text{s}}^h := \left\lbrace \bm s^h \Bigg| \begin{matrix}
%	\underline{\bm s}^h \leq \bm s^h \leq \overline{\bm s}^h, \\
%	- \bm b^h \leq \bm B^h \bm s^h \leq  \overline{ \bm q}^h -  \bm b^h \\
%	- \bm \epsilon + (1 - \bm a^T) \leq \bm a^T \bm s^h \leq \epsilon
%	\end{matrix}  \right\rbrace,
%\end{align}
%We define the following matrices and vectors: $\bm B^h = \col{(B_j)_{j=1}^{\nhs}}$ with lower triangular matrices  $[B_j^h]_{ik} = a_j^{i-k} \in \Rset{T \times T}$,  $ \overline{\bm q}^h = \col{(\overline{q}_j)_{j=1}^{\nhs}}$, vector $[b_j^h]_i = a^i$, $\bm b^h = \col{(b_jq_j(0))_{j=1}^{\nhs}}$ and $[a_j]_j = a^{T-i}$ with $a_j \in \Rset{T}$ and $\bm  a^h = \col{(a_j)_{j=1}^{\nhs}}$, $ \bm \epsilon = \col{(\epsilon_j)_{j=1}^{\nhs}}$. 
 
\section{Conclusion}\label{sec:conclusion}
In this work we provide an algorithm for solving constrained multi-cluster games and prove linear convergence to the optimum. As cooperative distributed optimization represents a special case of optimization in multi-cluster games, the presented algorithm extends existing literature on gradient-tracking methods proposed for unconstrained distributed optimization problems to the constraint ones. In the simulation part, we formulate a power management problem between microgrids as a multi-cluster problem and provided results on convergence time and discussions regarding the optimization result. In future work, we aim to extend our algorithm such that generalized Nash equilibrium problems, i.e. coupling constraints between the clusters, can be handled as well. 
 \vspace*{-\baselineskip}

\section*{Acknowledgments}
The work was gratefully supported by the German Research Foundation
(Deutsche Forschungsgemeinschaft, DFG) within the SPP 1984: Hybrid and multimodal energy systems: System theoretical methods for the transformation and operation of complex networks.

 \vspace*{-\baselineskip}
\appendix
\subsection{Supporting Proposition} \label{ap:supporting_propostion}

\begin{prop}\label{prop:Malpha}
	Let Assumptions~\ref{as:setX}, \ref{as:local_costfunction} \ref{as:mapping}, and \ref{as:global_graph} hold. Let 
	\begin{equation}\label{eq:Malpha}
	A(\alpha) = \begin{bmatrix}
	\frac{\alpha^2  L_0^2}{N} - \frac{2\alpha\mu}{N}  + 1      & \alpha^2 \frac{LL_0}{\sqrt N} + \alpha \left( L +   \frac{\sigma L_0}{\sqrt{N}} \right)\\
	\alpha^2 \frac{LL_0}{\sqrt N} + \alpha \left( L +   \frac{\sigma L_0}{\sqrt{N}} \right) & \alpha^2 L^2 + 2 \alpha \sigma^2 L + \sigma^2   
	\end{bmatrix}
	\end{equation}
	Then, with $\xstar = \1_N \otimes x^*$, the  inequality $
	\nl \xhat(k) - \xstar - \alpha \Qt \M(\x(k)) + \alpha \Qt \M(\xstar) \nr 
	\qquad \leq  \sqrt{\rho(A(\alpha))} \nl \x(k) - \xstar \nr 
	$
	holds. Furthermore, there exists a positive constant $\alpha_A$ dependent on the parameters $N$, $L$, $L_0$, $\mu$, and $\sigma$, such that $\sqrt{\rho(A(\alpha))}   < 1$ holds for any step size $\alpha\in(0,\alpha_A)$.
	% The upper bound $\alpha_A$ is further defined in the proof. 

\end{prop}
\begin{IEEEproof}
	The derivation of matrix $A(\alpha)$ and upper bounding the expression $||\xhat(k) - \xstar  - \alpha \Q^T \M(\x(k)) + \alpha \Q^T \M(\xstar)||$ by $\sqrt{\rho(A(\alpha))} ||\x(k) - \xstar ||$ is closely related to the proof of Theorem~1 of \cite{Bianchi2021} and shall therefore be omitted here due to space reasons.	In order to define the setting for $\alpha$ under which $\sqrt{\rho(A(\alpha))} < 1$, it can first be shown, by invoking Lemma~\ref{lemma:lipschitz_extended_mapping}, that all elements of the matrix $A(\alpha)$ are non-negative for $\alpha >0$. Next, it can be proven that the matrix $A(\alpha)$ is positive definite by demonstrating that this matrix is diagonal dominant, i.e. $a_{11}, a_{22} > a_{12} = a_{21}$, for  $0 < \alpha < \alpha_{A, 1}$. To do so, one needs to compare the corresponding quadratic polynomials in the elements of $A(\alpha)$. It is possible to explicitly specify the upper bound $\alpha_{A, 1}$ based on problem parameters, the details however shall be omitted here due to space limitation. Next, we apply Sylvester's criterion to find upper bounds on $\alpha$ which guarantees that the symmetric matrix $I_2 - A(\alpha)$ is positive definite. $ 1 - a_{11} > 0$ holds if $\alpha < \alpha_{A, 2}^1 = \frac{2\mu}{L_0^2}$. Next,  $\det(I_2 - A(\alpha))$ is positive if $q_1 \alpha^2 + q_2 \alpha - q_3 < 0$, where $q_1, q_2, q_3 >0$ are problem dependent factors (see the definition of $A(\alpha)$). Therefore, the determinant of the quadratic polynomial is positive and the inequality above holds, if $0 < \alpha < \alpha_{A, 2}^2 = \frac{-q_2 + \sqrt{q_2^2 + 4q_1q_3}}{2q_1} > 0$.
	Combining all above results, we have $\aoo  < 1$ if $0 < \alpha < \alpha_A = \min \lbrace \alpha_{A,1}, \alpha_{A,2} \rbrace$ with $\alpha_{A,2} = \min \lbrace \alpha_{A,2}^1, \alpha_{A,2}^2 \rbrace$ , which concludes the proof.
\end{IEEEproof}
\vspace*{-\baselineskip}
\subsection{Proof of Lemma \ref{lemma:equivalences}} \label{ap:proof_equivalences}
\begin{IEEEproof}
	Part $a) \iff b)$: 
	The equivalence between the solutions to variational inequalities and Nash equilibria in games is proven by Proposition~1.4.2 in \cite{Facchinei2007}. \\
%	Because $x^*$ is a solution of $\vi{\Omega, M}$, it is a fixed point of the mapping $H(x) = \text{proj}_{\Omega}\left[x - \alpha M(x)\right]$, i.e. $x^* = \text{proj}_{\Omega}\left[x^* - \alpha M(x^*)\right]$. In order for some $\xhopt$ to be a cooperative optimum inside cluster $h$ it needs to hold for some constant $x^{-h}$ that $\xhopt = \text{proj}_{\Omega^h}\left[\xhopt - \alpha \nabla_h F^h(\xhopt, x^{-h})\right]$. It can be observed that $\xhstar$, $\forall h \in \Hset$ are fixed points of the respective mappings if the combined vector $x^* = \col{(\xhstar)_{h \in \Hset}}$ is a fixed point of mapping $H(x)$. Therefore, if $x^*$ is the Nash equilibrium of the non-cooperative game between the clusters, the cooperative optima inside the clusters are achieved as well. \\
	Part $a) \iff c)$: 
	By definition $\xstar$ fulfils $(\I_{Nn} - \W)\xstar = 0$, i.e. describes a consensus state, in which all estimations are equal to $x^*$, which is a fixed point of $\Hb(\x)$ if $\Hb(\xstar) - \xstar = \bm 0$. Considering the definition of $\Q$ and Mapping $\M(\x)$ and by deleting identical and zero entries of the vector $\Hb(\xstar) - \xstar$, the fixed-point condition of mapping $\Hb(\x)$ at some consensus vector $\xstar$ reduces to $\col{(\projOmh{x^{h,*} - \alpha F^h(x^*)})_{h \in \Hset}} - \xhstar = 0$, which is equivalent to 
	$\text{proj}_{\Omega}\left[x^* - \alpha M(x^*)\right] - x^*$. Therefore $\xstar = 1_N \otimes x^*$ is a fixed-point of the mapping $\Hb(\x)$ if $x^*$ is a fixed-point of mapping $\text{proj}_{\Omega}\left[x^* - \alpha M(x^*)\right]$. The latter holds true if $x^*$ is a solution of $\vi{\Omega, M}$.\\
	Part $b) \iff d)$: 
	Let the estimations of the agents, stored in $\x$ be in consensus, i.e. it holds that $(\I_{Nn} - \W)\x = 0$. Then, the formulations of the problem in \eqref{prob:mcg} and \eqref{prob:mcg_extended} are exactly the same. As $(\I_{Nn} - \W)\xstar = 0$ holds and $x_i = x^*$ for $i =1, ..., N$, $\xstar$ is a solution of the extended problem in \eqref{prob:mcg_extended} if and only if $x^*$ is the solution of the central problem in \eqref{prob:mcg}. \\
	In conclusion $a) \iff b) \iff c) \iff d)$ holds. 
\end{IEEEproof}

\vspace*{-\baselineskip}

\subsection{Proof of Proposition \ref{prop:linear_inequality}} \label{ap:proof_linear_inequality}
\begin{IEEEproof}\\
	The proof is subdivided into two parts. In the first part, we show that the update equations can be expressed as a linear inequality system and show the structure of matrix $\Atau{\alpha}$. In the second part, we provide results on the spectral radius on $\alpha$ and define the upper bound $\overline{\alpha}$. \\
	\textbf{Part 1 - Linear system inequality:}
	 Using the fixed-point assumption $ \xstar = \projOm{\xstar - \alpha\Qt\M(\xstar)}$  and Equation~\eqref{alg:vectorwise_x}, we receive\vspace{-0.2cm}
	\begin{align*}
		& \nl \x(k+1)  - \xstar \nr \\
		&= \nl \projOm{\xhat(k)- \alpha \Qt \y(k)} - \projOm{\xstar- \alpha \Qt \M(\xstar)} \nr \\
		& \leq \nl \xhat(k)- \alpha \Qt \y(k) - \xstar + \alpha \Qt \M(\xstar) \nr \\
		& \leq \nl \xhat(k) - \xstar - \alpha \Qt \M(\x(k)) + \alpha \Qt \M(\xstar) \nr \\ 
		& \qquad + \nl  \alpha \Qt \M(\x(k)) - \alpha \Qt \y(k) \nr 
	\end{align*}
	For the first inequality, we used the non-expansiveness of the projection operator. Applying the results of Proposition~\ref{prop:Malpha} and 
		\begin{align*}
		&\nl  \alpha \Qt \M(\x(k)) - \alpha \Qt \y(k) \nrq \\
		&\leq \alpha^2 \nl \Qt \nrq \nl \M(\x(k)) - \y(k) \nrq \\
		& \leq \alpha^2 \nl \Qt \nrq \sum_{h=1}^H\nl \1_{N_h} \otimes \frac{1}{N_h} \sum_{i = 1}^{N_h} \nabla_h f_i^h(x_i(k)) - \y^h(k) \nrq \\
		& = \alpha^2 \sum_{h=1}^H \nl \y^h(k) - \bary^h(k) \nrq  = \alpha^2 \nl \y(k) - \bary(k) \nrq,
	\end{align*}
	where we used the definition of $\bary^h(k)$, we receive
	\begin{align*}
		\nl \x(k+1) - \xstar \nrq &\leq   \sqrt{\rho(A(\alpha))} \nl \x(k) - \xstar \nr \\ 
		&+ \alpha \nl \y(k) - \bary(k) \nr.
	\end{align*}
	Next, we consider the norm $\nl \y(k+1) -  \bary(k+1) \nr$ and the update equation \eqref{alg:vectorwise_y}:
	\begin{align*}
		& \nl \y(k+1) - \bary(k+1) \nr = || \V \y(k)+ \G(k+1) - \G(k) \\ 
		 & \qquad - \R\left( \y(k) + \G(k+1) - \G(k) \right)|| \\
		& \leq \nl  \V \y(k) - \bary(k) \nr +  \nl \bm I - \R \nr \nl \G(k+1) - \G(k) \nr
	\end{align*}
	Here, we used the double stochastic property of each $V^h$, $\forall h \in \Hset$ of Assumption~\ref{as:local_graphs}, i.e. the fact that $\R \V = \R$, which follows from the property $AC \otimes BD = (A\otimes B)(C \otimes D)$ of the Kronecker product, such that
	\begin{align*}
		\R^h \V^h &= \left( \frac{1}{N_h} \1_{N_h} \1^T_{N_h} \otimes \I_{n_h}\right)\left(V^h \otimes \I_{n_h}  \right) \\
		& = \left(\frac{1}{N_h} \1_{N_h} \1^T_{N_h} V^h \right) \otimes \left(\I_{n_h}\I_{n_h}\right) \\
		& = \left(\frac{1}{N_h} \1_{N_h} \1^T_{N_h} \right) \otimes  \I_{n_h} = \R^h,
	\end{align*}
	where the second to last equality is due to Assumption~\ref{as:local_graphs}. Using this result, we receive $\R\V = \diags{(\R^h\V^h)_{h \in \Hset}} = \R$.  
	Furthermore, using Lipschitz continuity of the local cost functions as discussed in Remark \ref{rem:lipschitz_continuity}, the inequalities
		\begin{align*}
		&\nl \G(k+1) - \G(k) \nrq = \sum_{h=1}^H \nl \G^h(k+1) - \G^h(k) \nrq \\
		&  = \sum_{h=1}^H \sum_{i = 1}^{N_h} \nl \nabla_h f_i^h(x_i(k+1)) - \nabla_h f_i^h(x_i(k)) \nrq \\
		& \leq L_1^2  \sum_{h=1}^H \sum_{i = 1}^{N_h} \nl x_i(k+1)\!-\! x_i(k) \nrq   = L_1^2 \nl \x(k+1) - \x(k) \nrq
	\end{align*}
	and 
	\begin{align*}
		&\nl \x(k+1) - \x(k)\nr  = \nl \projOm{\xhat(k) - \alpha \Qt\y(k)} - \x(k) \nr \\
		&= \nl \projOm{\xhat(k) - \alpha \Qt\y(k)} - \projOm{\xstar - \alpha\Qt\M(\xstar)} \nr \\ 
		& \qquad + \nl \x(k) - \xstar \nr \\
		& \leq \nl \xhat(k) - \alpha \Qt\y(k) - \xstar + \alpha\Qt\M(\xstar) \nr +  \nl \x(k) - \xstar \nr \\
		& \leq \nl \xhat(k) - \alpha \Qt  \M(\x(k))  - \xstar + \alpha\Qt\M(\xstar) \nr + \\ 
		& \quad \alpha \nl \Qt\M(\x(k)) - \Qt \y(k) \nr  +  \nl \x(k) - \xstar \nr \\
		& \leq (\sqspecAalpha +1) \nl \x(k) - \xstar \nr + \alpha \nl \y(k) - \bary(k) \nr
	\end{align*}
	hold, where we used the result of Proposition~\eqref{prop:Malpha} for the last inequality. Combining above results and applying Equation~\eqref{eq:sigmav}, we receive the following bound for the second norm
	\begin{align*}
		&\nl \y(k+1) - \bary(k+1) \!\nr \!\leq (\sigma_V \!+ \!\alpha L_1 \nl \bm I \!- \!\R \nr) \nl \y(k) - \bary(k) \nr \\ 
		&+ L_1(\sqspecAalpha +1) \nl \bm I - \R \nr \nl\x(k)- \xstar \nr
	\end{align*}
	The first part of the proof concludes with summarizing the elements of above inequalities in the matrix 
	\begin{equation}
	A_\tau(\alpha)= \begin{bmatrix}
	\aoo & \alpha \\
	a_{21}(\alpha) & \sigma_V + \alpha a_{22}
	\end{bmatrix},
	\end{equation}
	with  $\sigma_{V} \in (0,1)$ as  in Equation~\eqref{eq:sigmav} , $a_{21}(\alpha) = L_1 (\aoo +1)\nl \bm I- \R \nr$ and $a_{22} =  L_1 \nl \bm I- \R \nr > 0$. \\
	\textbf{Part 2 - Spectral radius:}
	According to Corollary~8.1.29 from \cite{Horn}, it holds for the spectral radius of $A_\tau(\alpha)$ that $\rho( A_\tau(\alpha)) \leq (1-\epsilon) < 1$ with some small $\epsilon > 0$,  if
	\begin{align}\label{eq:application_corollary}
		\begin{pmatrix}
			\aoo & \alpha \\
			a_{21}(\alpha) & \sigma_V + \alpha a_{22}
		\end{pmatrix} \begin{pmatrix}
			x_1 \\ x_2
		\end{pmatrix} \leq (1-\epsilon) \begin{pmatrix}
			x_1 \\ x_2
		\end{pmatrix}
	\end{align}  
	for some positive $x_1, x_2 > 0$.
	Now, we aim to find bounds on $\alpha$ such that condition above is satisfied. From Proposition~ \ref{prop:Malpha} we know that  $0 < \aoo < 1$ if $0 < \alpha < \alpha_A$, where $\alpha_A$ is some constant defined by the parameters of the game-theoretic problem.  Additionally, we provide the bound $0 < \alpha < (1- \sigma_V)/a_{22} = \alpha_{\sigma} >0 $. If this condition holds, then we can replace the term $\sigma_V + \alpha a_{22}$ by $\sigma_{v,\alpha}\in (0,1)$.  Consider
	\begin{equation*}
		\alpha \leq \frac{x_1}{x_2}  \left(1- \epsilon - \aoo\right), \frac{x_1}{x_2} \leq  \frac{(1- \epsilon - \sigma_{v,\alpha})}{a_{21}(\alpha)},
	\end{equation*}
	which follows from the inequality system of Equation~\eqref{eq:application_corollary}. Combining the two inequalities results in $
		\alpha \leq  \frac{ (1- \epsilon - \aoo)(1- \epsilon - \sigma_{v,\alpha})}{a_{21}(\alpha)} = \alpha_\epsilon.$
	Note that $a_{21}(\alpha) > 0$.  With $\alpha^*(\alpha) = \lim_{\epsilon \rightarrow 0} \alpha_\epsilon = \frac{ (1 - \aoo)(1 - \sigma_{v,\alpha})}{a_{21}(\alpha)} $, choosing $
	\alpha \in(0, \min\lbrace \alpha^*(\alpha), \alpha_A, \alpha_{\sigma} \rbrace) $
	results in 	$\rho(A_\tau(\alpha)) < 1$. Next we demonstrate how to set up $\alpha$ satisfying $
	0 < \alpha < \min\lbrace \alpha^*(\alpha), \alpha_A, \alpha_{\sigma} \rbrace $. \\
	Let us note that for any $\alpha >0$, such that $\alpha <  \min\lbrace \alpha_A/2, \alpha_{\sigma} \rbrace = \alpha_{A, \sigma}$, we have $\alpha^*(\alpha) >  \frac{ \delta_1(\alpha) \delta_2(\alpha)}{	2 L_1 ||I - \R|| } $ with $\delta_1(\alpha) = 1 - \aoo > 0$ and $\delta_2(\alpha) = 1 - \sigma_{v,\alpha} > 0$. By leveraging Corollary~8.1.19 from \cite{Horn} and investigating matrix $A(\alpha)$ of Proposition~\ref{prop:Malpha}, we find that $\sqrt{\rho(A(\alpha))} \leq \sqrt{\rho(A(\alpha_{A, \sigma}))}$, wherefore $\delta_1(\alpha) \geq \delta_1(\alpha_{A, \sigma})$ and $\delta_2(\alpha) > \delta_2(\alpha_{A, \sigma})$. In conclusion, if $\alpha < \alpha_{A, \sigma}$, then $\alpha^*(\alpha) > \alpha^*(\alpha_{A, \sigma})>0$. Therefore, choosing $0 < \alpha <\overline{\alpha} = \min\lbrace \alpha^*(\alpha_{A, \sigma}), \alpha_{A, \sigma} \rbrace$ with $\alpha_{A, \sigma} = \min \lbrace \alpha_A, \alpha_{\sigma} \rbrace $ ensures $\rho(\Atau{\alpha}) < 1$. 
\end{IEEEproof}

\bibliographystyle{IEEEtran}

\begin{thebibliography}{10}
	\providecommand{\url}[1]{#1}
	\csname url@samestyle\endcsname
	\providecommand{\newblock}{\relax}
	\providecommand{\bibinfo}[2]{#2}
	\providecommand{\BIBentrySTDinterwordspacing}{\spaceskip=0pt\relax}
	\providecommand{\BIBentryALTinterwordstretchfactor}{4}
	\providecommand{\BIBentryALTinterwordspacing}{\spaceskip=\fontdimen2\font plus
		\BIBentryALTinterwordstretchfactor\fontdimen3\font minus
		\fontdimen4\font\relax}
	\providecommand{\BIBforeignlanguage}[2]{{%
			\expandafter\ifx\csname l@#1\endcsname\relax
			\typeout{** WARNING: IEEEtran.bst: No hyphenation pattern has been}%
			\typeout{** loaded for the language `#1'. Using the pattern for}%
			\typeout{** the default language instead.}%
			\else
			\language=\csname l@#1\endcsname
			\fi
			#2}}
	\providecommand{\BIBdecl}{\relax}
	\BIBdecl
	
	\bibitem{Bianchi2020}
	M.~Bianchi and S.~Grammatico, ``{A continuous-time distributed generalized Nash
		equilibrium seeking algorithm over networks for double-integrator agents},''
	\emph{European Control Conference 2020, ECC 2020}, pp. 1474--1479, may 2020.
	
	\bibitem{Bianchi2021}
	------, ``{Fully Distributed Nash Equilibrium Seeking over Time-Varying
		Communication Networks with Linear Convergence Rate},'' \emph{IEEE Control
		Systems Letters}, vol.~5, no.~2, pp. 499--504, apr 2021.
	
	\bibitem{Pavel2020}
	L.~Pavel, ``{Distributed GNE Seeking under Partial-Decision Information over
		Networks via a Doubly-Augmented Operator Splitting Approach},'' \emph{IEEE
		Transactions on Automatic Control}, vol.~65, no.~4, pp. 1584--1597, apr 2020.
	
	\bibitem{Zimmermann2020}
	J.~Zimmermann, T.~Tatarenko, V.~Willert, and J.~Adamy, ``{Projected Push-Sum
		Gradient Descent-Ascent for Convex Optimization with Application to Economic
		Dispatch Problems},'' \emph{IEEE Conference on Decision and Control}, vol.
	2020-December, pp. 542--547, dec 2020.
	
	\bibitem{Tatarenko2019}
	T.~Tatarenko, J.~Zimmermann, V.~Willert, and J.~Adamy, ``{Penalized Push-Sum
		Algorithm for Constrained Distributed Optimization with Application to Energy
		Management in Smart Grid},'' in \emph{IEEE Conference on Decision and
		Control}, vol. 2019-Decem, dec 2019, pp. 6234--6241.
	
	\bibitem{Chen2018}
	G.~Chen and Q.~Yang, ``{An ADMM-Based Distributed Algorithm for Economic
		Dispatch in Islanded Microgrids},'' \emph{IEEE Transactions on Industrial
		Informatics}, vol.~14, no.~9, pp. 3892--3903, sep 2018.
	
	\bibitem{Zhao17}
	C.~Zhao, J.~He, P.~Cheng, and J.~Chen, ``Consensus-based energy management in
	smart grid with transmission losses and directed communication,'' \emph{IEEE
		Transactions on Smart Grid}, vol.~8, no.~5, pp. 2049--2061, 2017.
	
	\bibitem{Wang2017}
	Y.~Wang, L.~Wu, and S.~Wang, ``A fully-decentralized consensus-based admm
	approach for dc-opf with demand response,'' \emph{IEEE Transactions on Smart
		Grid}, vol.~8, no.~6, pp. 2637--2647, 2017.
	
	\bibitem{Belgioioso2020}
	G.~Belgioioso, W.~Ananduta, S.~Grammatico, and C.~Ocampo-Martinez, ``{Energy
		Management and Peer-to-peer Trading in Future Smart Grids: A Distributed
		Game-Theoretic Approach},'' \emph{European Control Conference 2020}, pp.
	1324--1329, may 2020.
	
	\bibitem{Gabriel2013}
	S.~A. Gabriel, A.~J. Conejo, J.~D. Fuller, B.~F. Hobbs, and C.~Ruiz,
	\emph{Complementarity Modeling in Energy Markets}.\hskip 1em plus 0.5em minus
	0.4em\relax Springer, 2013.
	
	\bibitem{Ye2020}
	M.~Ye, G.~Hu, and S.~Xu, ``{An extremum seeking-based approach for Nash
		equilibrium seeking in N-cluster noncooperative games},'' \emph{Automatica},
	vol. 114, p. 108815, 2020.
	
	\bibitem{Sun2021}
	C.~Sun and G.~Hu, ``{Distributed Generalized Nash Equilibrium Seeking of
		N-Coalition Games with Full and Distributive Constraints},'' \emph{arXiv
		2109.12515}, sep 2021.
	
	\bibitem{Tatarenko2021}
	T.~Tatarenko, J.~Zimmermann, and J.~Adamy, ``Gradient play in n-cluster games
	with zero-order information,'' pp. 3104--3109, 2021.
	
	\bibitem{Qu2018}
	G.~Qu and N.~Li, ``{Harnessing smoothness to accelerate distributed
		optimization},'' \emph{IEEE Transactions on Control of Network Systems},
	vol.~5, no.~3, pp. 1245--1260, sep 2018.
	
	\bibitem{Pu2021}
	S.~Pu, W.~Shi, J.~Xu, and A.~Nedic, ``{Push-Pull Gradient Methods for
		Distributed Optimization in Networks},'' \emph{IEEE Transactions on Automatic
		Control}, vol.~66, no.~1, pp. 1--16, jan 2021.
	
	\bibitem{Dong2020}
	Z.~Dong, S.~Mao, W.~Du, and Y.~Tang, ``{Distributed Constrained Optimization
		with Linear Convergence Rate},'' in \emph{IEEE International Conference on
		Control and Automation}, vol. 2020-October, oct 2020, pp. 937--942.
	
	\bibitem{Falsone2022}
	A.~Falsone and M.~Prandini, ``{Distributed decision-coupled constrained
		optimization via Proximal-Tracking},'' \emph{Automatica}, vol. 135, p.
	109938, jan 2022.
	
	\bibitem{Meng2020}
	M.~Meng and X.~Li, ``{On the linear convergence of distributed Nash equilibrium
		seeking for multi-cluster games under partial-decision information},''
	\emph{arXiv 2005.06923}, 2020.
	
	\bibitem{Zimmermann2021}
	J.~Zimmermann, T.~Tatarenko, V.~Willert, and J.~Adamy, ``{Solving leaderless
		multi-cluster games over directed graphs},'' \emph{European Journal of
		Control}, vol.~62, pp. 14--21, nov 2021.
	
	\bibitem{Pang2020}
	Y.~Pang and G.~Hu, ``{Fully Distributed Nash Equilibrium Seeking in N-Cluster
		Games},'' \emph{arXiv 2012.11347}, dec 2020.
	
	\bibitem{Zhou2021}
	J.~Zhou, Y.~Lv, G.~Wen, J.~Lv, and D.~Zheng, ``{Distributed Nash Equilibrium
		Seeking in Consistency-Constrained Multi-Coalition Games},'' \emph{arXiv
		2106.10513}, jun 2021.
	
	\bibitem{TatShiNedich21}
	T.~Tatarenko, W.~Shi, and A.~Nedi\'c, ``Geometric convergence of gradient play
	algorithms for distributed nash equilibrium seeking,'' \emph{IEEE
		Transactions on Automatic Control}, vol.~66, no.~11, pp. 5342--5353, 2021.
	
	\bibitem{Facchinei2007}
	F.~Facchinei and J.-S. Pang, \emph{Finite-dimensional variational inequalities
		and complementarity problems}.\hskip 1em plus 0.5em minus 0.4em\relax
	Springer Science \& Business Media, 2007.
	
	\bibitem{Horn}
	R.~A. Horn and C.~R. Johnson, \emph{{Matrix Analysis}}.\hskip 1em plus 0.5em
	minus 0.4em\relax Cambridge University Press, 1990.
	
\end{thebibliography}

\newpage

\vfill

\end{document}